\documentclass[aps,prl,showpacs,floatfix,groupedaddress,twocolumn]{revtex4-1}
\usepackage{bm,color,amsmath,amssymb,siunitx,blindtext}
\usepackage{graphicx,epstopdf}

\usepackage[latin1]{inputenc}
\newcommand{\beq}{\begin{equation}}
\newcommand{\eeq}{\end{equation}}
\newcommand{\bea}{\begin{eqnarray}}
\newcommand{\eea}{\end{eqnarray}}

\newcommand{\lin}{\mathrm{lin}}
\newcommand{\ind}{\mathrm{ind}}

\begin{document} \title{Generalized Euler-Lotka equation for correlated cell divisions}  

 \author{Simone Pigolotti} \email{simone.pigolotti@oist.jp} 
\affiliation{Biological Complexity Unit, Okinawa Institute of Science and Technology and Graduate University, Onna, Okinawa 904-0495.}  

\begin{abstract} Cell division times in microbial populations display significant fluctuations, that impact the population growth rate in a non-trivial way. If fluctuations are uncorrelated among different cells, the population growth rate is predicted by the Euler-Lotka equation, which is a classic result in mathematical biology. However, cell division times can be significantly correlated, due to physical properties of cells that are passed through generations.  In this paper, we derive an equation remarkably similar to the Euler-Lotka equation which is valid in the presence of correlations. Our exact result is based on large deviation theory and does not require particularly strong assumptions on the underlying dynamics. We apply our theory to a phenomenological model of bacterial cell division in E.coli and to experimental data. We find that the discrepancy between the growth rate predicted by the Euler-Lotka equation and our generalized version is relatively small, but large enough to be measurable by our approach. 
\end{abstract}


\maketitle

Microbial populations in steady, nutrient-rich conditions tend to grow exponentially. Their exponential growth rate $\Lambda$ can be taken as a proxy for the population fitness and is therefore a biologically important quantity. In a population of cells dividing at regular times $\tau$, the population size at a time $T$ multiple of $\tau$ is  $N(T)=N(0)2^{T/\tau}$, so that $\Lambda=\ln 2/\tau$.  In practice, cell division times of microbial populations significantly fluctuate, so that the division time $\tau_i$ of a given individual $i$ must be considered as a random quantity. As a consequence, the growth of $N(T)$ is stochastic. In these situations, we can still define an exponential growth rate by
\begin{equation}\label{eq:lambda}
\Lambda=\lim_{T\rightarrow \infty} \frac{1}{T}\ln N(T) .
\end{equation}
For independent, identically distributed cell division times $\tau_i$, the exponential growth rate converges to a deterministic value and can be computed as solution of the celebrated Euler-Lotka equation
\begin{equation}\label{eq:EL}
2\,\langle e^{-\Lambda \tau}\rangle_{\tau}=1, 
\end{equation}
where we denote the average over the distribution $p(\tau)$ of the division times by $\langle f(\tau) \rangle_{\tau}=\int d\tau\, p(\tau) f(\tau)$. We use this notation also for discrete variables, with the integral appropriately replaced by a sum. Equation~\eqref{eq:EL} is a classic result in mathematical biology. A recent experimental study has tested its prediction by tracking individual cell divisions in a microfluidic device \cite{hashimoto2016noise}. Beside microbial populations, the Euler-Lotka equation finds important applications in epidemiology, where the factor $2$ is replaced by the reproductive number $R_0$ \cite{grassly2008mathematical}. 

Experimental studies have revealed that fluctuations in microbial cell features are correlated among generations \cite{wang2010robust,campos2014constant,wallden2016synchronization}. These correlations are caused by properties of cells that are passed from mother to daughter. These properties can be physical such as cell mass, or biological such as gene expression. Their fluctuations are often controlled to preserve homeostasis, i.e., a stable state of cells across generations. For example, experimental and theoretical studies provided evidences for an "adder" mechanism, in which cells attempt at growing their mass by a constant amount before dividing \cite{amir2014cell,campos2014constant,taheri2015cell}.  

Regardless of the underlying mechanism, generalizing Eq.~\eqref{eq:EL} to correlated cell divisions has proven to be a hard problem. One relatively simple case is the ``Markovian'' scenario where a cell division time conditionally depends only on that of her mother. Expressions for the growth rate in these cases have been derived in classic works by Powell \cite{powell1956growth} and Lebowitz and Rubinow \cite{lebowitz1974theory}, see also \cite{lin2017effects,levien2020interplay}. Alternative approaches estimate the growth rate by comparing the outcome of sampling the population forward in time with retrospective sampling, in which individuals in the final population are traced back to their ancestors \cite{kobayashi2015fluctuation,garcia2019linking,genthon2020fluctuation,genthon2020universal}. A recent study links the exponential growth rate $\Lambda$ to the asymptotic distribution of the number of cell divisions $\Delta$ among lineages \cite{levien2020large} using large deviation theory. This approach has the advantage of neither requiring the Markovian assumption, nor retrospective sampling.

In this paper, we introduce a generalized Euler-Lotka equation~\eqref{eq:EL} which is valid for correlated cell division times:
\begin{equation}\label{eq:generalizedEL}
\lim_{\Delta\rightarrow \infty}\frac{1}{\Delta}\ln \left\langle e^{-\Lambda \sum\limits_{i=1}^{\Delta}{\tau_i}}\right\rangle_{\{\tau_i\}}=-\ln 2 ,
\end{equation} 
where $\langle \dots \rangle_{\{\tau_i\}}$ denotes an average over sequences $\{\tau_i \}=\tau_1,\tau_2\dots \tau_{\Delta}$ of cell division times in independent lineages. If the $\tau_i$s were uncorrelated, then the left hand side of Eq.~\eqref{eq:generalizedEL} reduces to the cumulant generating function $\ln \langle e^{q\tau}\rangle_{\tau}$, and therefore Eq.~\eqref{eq:generalizedEL} becomes equivalent to the traditional Euler-Lotka equation \eqref{eq:EL}. Equation~\eqref{eq:generalizedEL} only requires as hypotheses that population dynamics is at steady state and the sum of the $\tau_i$s across a lineage satisfies a large deviation principle with a convex rate function, which in practice are rather mild assumptions (see, e.g., sections 3.5.6 and 4.4 of Ref.~\cite{touchette2009large}). Equation~\eqref{eq:generalizedEL} can therefore be used to compute the population growth rate from individual lineages in rather general settings.

\begin{figure}[htb]
\begin{center}
\includegraphics[width=8cm]{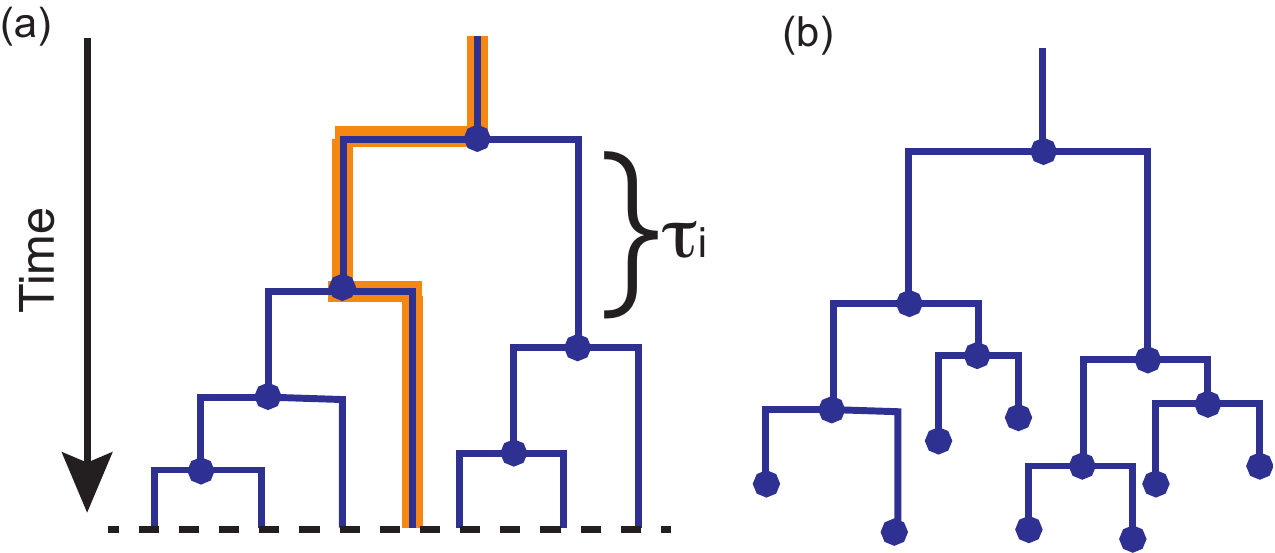}
\caption{Population dynamics represented as a lineage tree. (a) a microbial population grows in time from a single cell. Nodes (circle) denote cell division events. Lengths of branches denote the cell division times $\tau_i$s. One lineage is represented with a thick orange line. In this case, the population is let to evolve until a fixed time $T$. (b) Lineage tree in an alternative ensemble, in which lineages are let to evolve until they have accumulated exactly $\Delta=4$ cell divisions.\label{fig:illust}}
\end{center}
\end{figure}

We consider a microbial population initially constituted of a single individual. The population grows in time by a sequence of cell divisions. We represent the genealogy of the population by a tree, whose nodes are cell division events and branches are times between consecutive cell divisions, see Fig.~\ref{fig:illust}a.  

We now introduce the concept of a lineage. A lineage is identified by an individual in the population at time $T$ complemented by its past history, i.e., the number $\Delta$ of cell divisions separating it from the individual at time $0$ and the sequence of cell division times $\{\tau_i\}$ of all its ancestors, see Fig.~\ref{fig:illust}a. Following Refs. \cite{nozoe2017inferring,levien2020large}, we now imagine to randomly select a lineage by starting from the initial individual and picking at each divisions one of the two newborns with equal probability. With this procedure, a lineage that includes $\Delta$ cell divisions is chosen with probability $2^{-\Delta}$. We approximate the probability that this randomly selected lineage includes $\Delta$ cell division events by the empirical frequency $p(\Delta;T)\approx 2^{-\Delta}N(\Delta;T)$, where $N(\Delta;T)$ is the number of lineages with $\Delta$ cell divisions at time $T$. Since $N(T)=\sum_\Delta N(\Delta;T)$, we obtain
\begin{equation}\label{eq:amir_nt}
N(T)\approx \langle 2^\Delta\rangle_\Delta.
\end{equation}
Substituting this expression into the definition of the exponential growth rate, Eq.~\eqref{eq:lambda}, we find
\begin{equation}\label{eq:amir_lambda}
\Lambda=\lim_{T\rightarrow \infty} \frac{1}{T}\ln \langle 2^\Delta\rangle_\Delta.
\end{equation}
At variance with Eq.~\eqref{eq:amir_nt}, Eq.~\eqref{eq:amir_lambda} is an exact equality, as the empirical frequencies $2^{-\Delta}N(\Delta;T)$ converge to $p(\Delta;T)$ in the limit $T\rightarrow \infty$ \cite{jafarpour2018bridging,levien2020large}.

To make further progress, we introduce some ideas from large deviation theory \cite{touchette2009large}. Large deviation theory describes the leading behavior of probability distributions when a parameter (like the time $T$ in our case) becomes large. In large deviation theory, variables such as $\Delta$, whose average is proportional to $T$, are called extensive. We associate with $\Delta$ the intensive variable $\delta=\Delta/T$, whose average tends to a constant for large $T$. The large deviation principle for $\delta$ is expressed by 
\begin{equation}
p(\delta) \asymp e^{-TI^{(\delta)}(\delta)}.
\end{equation}
The function $I^{(\delta)}(\delta)$ is called the rate function. We use the notation $I^{(\delta)}$ to stress that $I$ is the rate function associated with the distribution of the variable $\delta$. The symbol ``$\asymp$'' denotes the leading exponential behavior; it can be seen as a shorthand for $I(\delta)=-\lim_{T\rightarrow\infty} [\ln p(\delta)]/T$. 

An alternative way of studying asymptotic fluctuations of intensive random variables is via the scaled cumulant generating function, defined by
\begin{equation}\label{eq:scaledcumul}
\psi^{(\delta)}(q)=\lim_{T\rightarrow \infty}\frac{1}{T}\ln \langle e^{qT\delta}\rangle_{\delta}.
\end{equation} 
The Gartner-Ellis theorem states that, if the rate function is convex, it is related with the scaled cumulant generating by a Legendre-Fenchel transform
\begin{equation}
I^{(\delta)}(\delta)=\sup_q \,[q\delta-\psi^{(\delta)}(q)] .
\end{equation}
Since the Legendre-Fenchel transform is an involution, it also holds that $\psi^{(\delta)}(q)=\sup_\delta \,[q\delta-I^{(\delta)}(\delta)]$.

We now return to Eq.~\eqref{eq:amir_nt} and briefly summarize the main result of Ref.~\cite{levien2020large}. Assuming that $\delta$ satisfies a large deviation principle, we obtain
\begin{equation}
\Lambda=\lim_{T\rightarrow \infty} \frac{1}{T}\ln \int d\delta e^{T[\delta \ln 2 -I^{(\delta)}(\delta)]}.
\end{equation}
In the limit $T\rightarrow \infty$, the integral can be evaluated with the method of steepest descent, obtaining
\begin{equation}\label{eq:amir_central}
\Lambda=\sup_{\delta} [\delta \ln 2 -I^{(\delta)}(\delta)].
\end{equation}
Equation~\eqref{eq:amir_central} is the central result of Ref.~\cite{levien2020large}. An alternative way to obtain it is to directly identify the expression of the scaled cumulant generating function in Eq.~\eqref{eq:amir_lambda}:
\begin{equation}\label{eq:psidelta}
\Lambda=\psi^{(\delta)}(\ln 2) .
\end{equation}
Equation~\eqref{eq:amir_central} then follows by expressing the scaled cumulant generating function in terms of the rate function by means of the Gartner-Ellis theorem.

Application of this theory requires knowledge of the asymptotic distribution of $\Delta$, or its intensive counterpart $\delta$. However, in analogy with the Euler-Lotka equation~\eqref{eq:EL} it would be desirable to express $\Lambda$ in terms of the distribution of division times and its correlations. To this aim, we consider a case in which, rather than letting the population grow until a given time $T$, each lineage is let to grow until it has accumulated exactly $\Delta$ cell divisions, see Fig.~\ref{fig:illust}b. In this alternative {\em ensemble}, $\Delta$ is fixed whereas $T$ fluctuates among lineages. In this case, we consider $T$ as an extensive random variable, since its average grows linearly with the fixed large parameter $\Delta$. We similarly associate with $T$ the intensive variable $t=T/\Delta$.  We expect $t$ to satisfy as well a large deviation principle:
\begin{equation}
p(t)\asymp e^{-\Delta I^{(t)}(t)}.
\end{equation}
In the language of probability theory and in particular of queuing theory, $\Delta(T)$ is called a counting process and $T(\Delta)$ its inverse. A useful result \cite{glynn1994large} states that the large deviation of their associated intensive variables, $\delta$ and $t$ respectively, are related by
\begin{equation}
I^{(\delta)}(x)=x I^{(t)}(1/x).
\end{equation}
Note that this is the result that one would obtain by simply applying the rule for a change of variable in the large deviation form of the probability distribution. 

We now substitute this result into Eq.~\eqref{eq:amir_central}, obtaining
\begin{equation}\label{eq:step}
\Lambda=\sup_{\delta} \left\{\delta \left[\ln 2 - I^{(t)}\left(\frac{1}{\delta}\right)\right]\right\},
\end{equation}
and, by applying the Gartner-Ellis theorem,
\begin{align}\label{eq:step2}
\Lambda&=\sup_{\delta} \left\{\delta \left[\ln 2 - \sup_q\left(\frac{q}{\delta}-\psi^{(t)}(q)\right)\right]\right\}\nonumber\\
&=\sup_{\delta}\inf_q \left[\delta \ln 2 -q+\delta\psi^{(t)}(q)\right] .
\end{align}
We assume that the function in square brackets smoothly depends on $\delta$ and $q$ and therefore compute the supremum and infimum by simply taking derivatives. The extremality condition respect to $\delta$ is expressed by
\begin{equation}\label{eq:EL_generalized_prel}
\psi^{(t)}(q^{\inf})=-\ln 2 .
\end{equation}
Substituting this condition back into Eq.~\eqref{eq:step2} yields $\Lambda=-q^{\inf}$, so that we rewrite Eq.~\eqref{eq:EL_generalized_prel} as
\begin{equation}\label{eq:EL_generalized}
\psi^{(t)}(-\Lambda)=-\ln 2 .
\end{equation}
Upon substituting the definition of the scaled cumulant generating function, Eq.~\eqref{eq:scaledcumul}, into Eq.~\eqref{eq:EL_generalized}, we obtain the generalized Euler-Lotka equation~\eqref{eq:generalizedEL}, as anticipated. 

Taking the derivative in Eq.~\eqref{eq:step2} respect to $q$ results in
\begin{equation}\label{eq:linkensembles}
\left.\delta^{\max}\frac{d}{dq}\psi^{(t)}(q)\right|_{q=-\Lambda}=1. 
\end{equation}
This equation relates the dominant value of $\delta$ with the statistics of the division times and provides another facet to the generalized Euler-Lotka theory. Equation~\eqref{eq:linkensembles} is best interpreted in the simple case of uncorrelated cell divisions, where it reduces to $\delta^{\max}=\langle \tau e^{-\Lambda \tau}\rangle^{-1}$. If $\Lambda\ll 1$, the dominant value of $\gamma$ is simply its average value, i.e. the inverse of the average division time. However, for quickly growing population, the dominant value of $\delta$ becomes significantly larger than this value, as cells that reproduce faster contribute more to population growth.

To illustrate our result, we consider a phenomenological model of bacterial growth inspired by the adder principle \cite{campos2014constant}. In the model, each bacterial cell grows in length at a rate $\alpha$. The rate $\alpha$ is generated by the formula
\begin{equation}
\alpha=\alpha_0 + c(\alpha_M-\alpha_0) +\sigma_\alpha (1-c^2) \xi .
\end{equation}
where $\alpha_M$ is the value of $\alpha$ of the mother of the considered cell. The parameters $\alpha_0$ and $\sigma_\alpha$ are the average and variance of the distribution of $\alpha$, respectively. The variable $\xi$ is a Gaussian random variable with zero average and unit variance. The parameter $c$ controls the degree of correlations between the growth rate of mothers and daughters. 

Each cell is characterized by a length $s_{b}$ at birth and $s_{d}$ at death. The adder model postulates that the added length $l=s_d-s_{b}$ is roughly constant among cells. After division, a daughter inherits a fraction $f$ of the mother's length. We allow for some variability by taking both $f$ and $l$ as random variables, whose distributions are estimated from experimental data \cite{campos2014constant}. The time between cell divisions is expressed by 
\begin{equation}\label{eq:adder}
\tau=\frac{1}{\alpha \ln(s_{d}/s_{b})}=\frac{1}{\alpha \ln(1+l/s_{b})}.
\end{equation}

\begin{figure}[htb]
\begin{center}
\includegraphics[width=9cm]{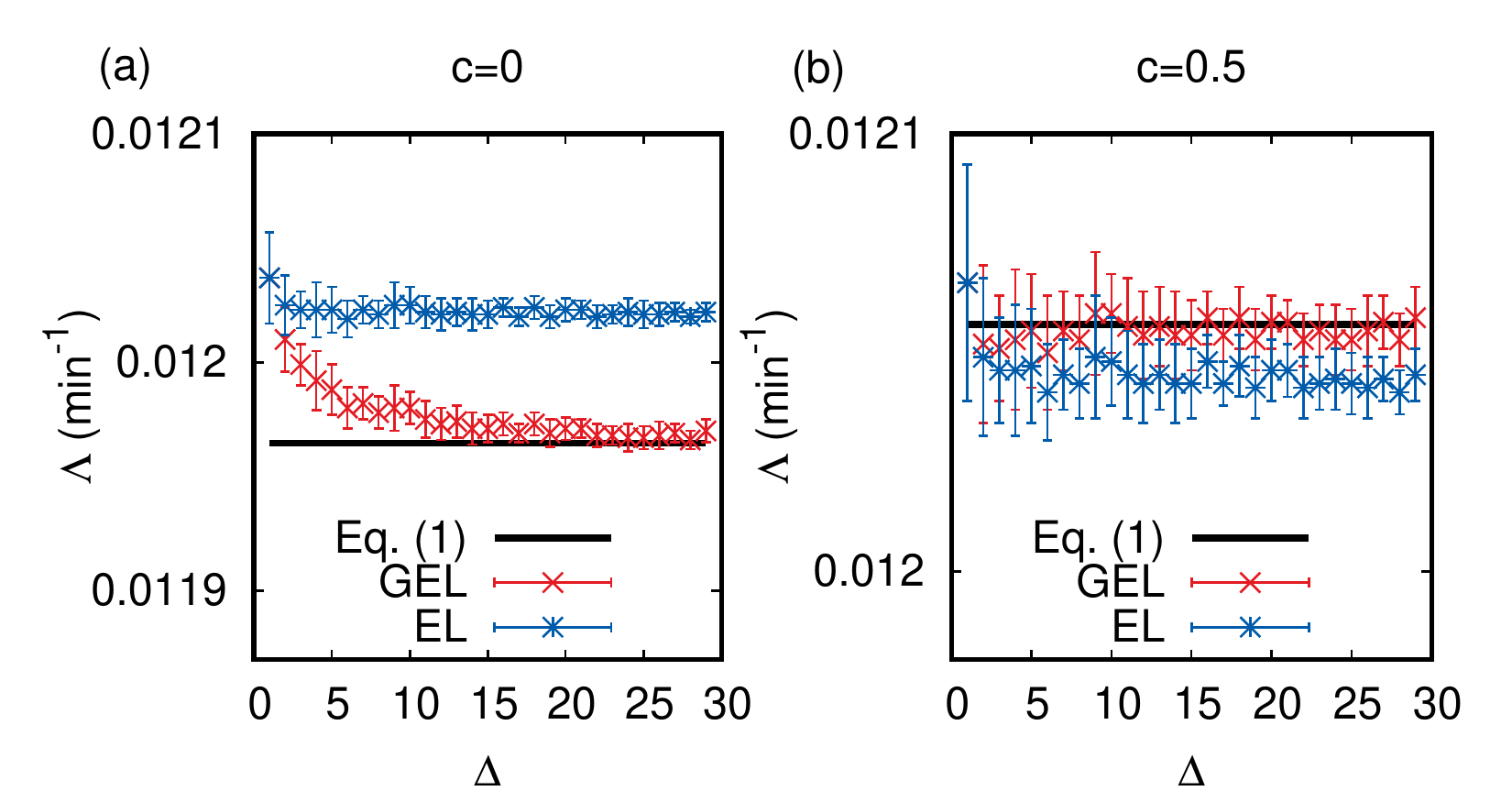}
\caption{Estimates of the growth rate $\Lambda$ in the adder model of Ref.~\cite{campos2014constant} obtained by the direct definition given in Eq.~\eqref{eq:lambda}, by the generalized Euler-Lotka equation~\eqref{eq:generalizedEL} (GEL), and by the conventional Euler-Lotka equation~\eqref{eq:EL} (EL).  In both panels, parameters of the growth rate distribution are $\alpha_0=0.0255\, \min^{-1}$ and $\sigma_\alpha=0.0027\, \min^{-1}$. The inherited length fraction $f$ is distributed according to a Gaussian with mean $f_0=0.5$ and standard deviation $\sigma_f=0.03$. The added length $l$ follows a lognormal distribution with mean $l_0=3.21\, \mu m$ and standard deviation $\sigma_l=0.54\, \mu m$. In panel (a), growth rates of mothers and daughters are uncorrelated ($c=0$). In panel (b) growth rates of mothers and daughters are positively correlated ($c=0.5$).  In both panels, we average over $n_{\lin}=5000$ lineages and plot the results as a function of the number of cell divisions $\Delta$ in each lineage. In all panels, simulations are repeated $20$ times; error bars denote standard deviations computed from these realizations. Details on the direct numerical estimate of $\Lambda$ are presented in the Appendix {\em Estimate of the exponential growth rate}.
\label{fig:adderfigure}}
\end{center}
\end{figure}

To understand the dynamics of this model, we first consider the simple case in which the growth rate $\alpha$ is uncorrelated across generation, $c=0$. In this case, the adder dynamics causes {\em negative} correlations in cell division times. An intuitive explanation is that cells that grow to a large size tend to have a long generation time and to give birth to larger cells. These larger cells, in turn tend to have a short generation time because of the dependence of $\tau$ on $s_{b}$ in Eq.~\eqref{eq:adder}. Since the traditional Euler-Lotka equation neglects correlations, we expect it to overestimate $\Lambda$ in this case. This idea is confirmed in Fig.~\ref{fig:adderfigure}a, which also shows how the generalized Euler-Lotka equation correctly estimates the directly measured value of $\Lambda$. 

We now move to a case in which growth rates are correlated across generations, $c=0.5$. This value is compatible with observation from E.coli experiments \cite{campos2014constant}. In this case, the positive correlations induced by the growth rate tend to counterbalance those caused by the adder model. As a result, both the traditional and the generalized Euler-Lotka equations predict a growth rate that is close to the correct one, although the traditional Euler-Lotka equation slightly underestimates it, see Fig.~\ref{fig:adderfigure}b. 

\begin{figure}[htb]
\begin{center}
\includegraphics[width=9cm]{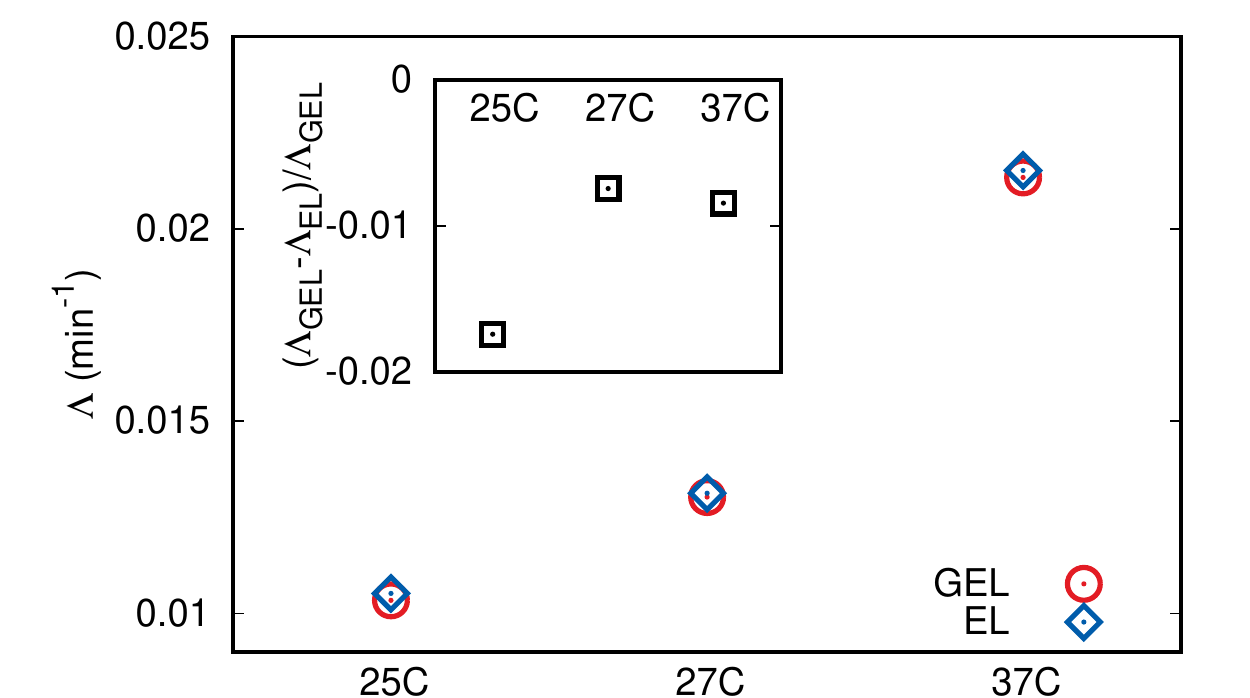}
\caption{Growth rate $\Lambda$ estimated from Eq.~\eqref{eq:EL} and Eq.~\eqref{eq:EL_generalized} from the experiments in \cite{tanouchi2015noisy,tanouchi2017long}. The experimental data consists of a number of lineages equal to $65$, $54$, $160$ for three different temperatures $25 \mathrm{C}$, $27 \mathrm{C}$, and $37 \mathrm{C}$, respectively. Each of these lineages includes $70$ division events. For the EL estimate, we consider all division events at each temperature. For the GEL estimate, we use all sub-lineages of length $\Delta=20$ from all lineages at each temperature. The inset shows the relative difference between the two estimates.
\label{fig:experiments}}
\end{center}
\end{figure}

We now test the generalized Euler-Lotka equation on experimental data from \cite{tanouchi2015noisy,tanouchi2017long}. These experiments tracked a large number of E.coli lineages at different temperatures. Applying our approach, we find that the generalized Euler-Lotka equation predicts a lower growth rate than the traditional Euler-Lotka equation, see Fig.~\ref{fig:experiments}. The differences are on the order of $1\%$ and suggest that the compensatory mechanisms observed in Ref.~\cite{campos2014constant} are less effective in the  experiments by \cite{tanouchi2015noisy,tanouchi2017long}. A direct inspection of the correlation functions of the cell division times support this idea, see Appendix {\em Correlation functions}.

An advantage of our approach is that it allows us to use the arsenal of techniques from large deviation theory \cite{touchette2009large} to compute the scaled cumulant generating function, and thereby the growth rate via Eq.~\eqref{eq:generalizedEL}. For example, we apply our theory to the Markovian case in which the division time conditionally depends on the maternal division time only, i.e.,
\begin{equation}
p(\tau_i|\tau_{i-1}, \tau_{i-2},\tau_{i-3}\dots)=p(\tau_i|\tau_{i-1}).
\end{equation}
In this case, one has
\begin{equation}\label{eq:convol}
\left\langle e^{q\sum_{i=1}^\Delta \tau_i}\right\rangle = \int d\tau_1\dots d\tau_\Delta\, p_{\tau_1}e^{q\tau_{\Delta}} \prod_{i=2}^\Delta e^{q\tau_{i-1}} p(\tau_i|\tau_{i-1}) .
\end{equation}
Equation~\eqref{eq:convol} expresses the generating function as an iterated convolution. This iterated convolution is characterized by the integral kernel $K(\tau_i,\tau_{i-1})=e^{q\tau_{i-1}} p(\tau_i|\tau_{i-1})$. For large $\Delta$, the behavior of the left-hand side of Eq.~\eqref{eq:convol} is dominated by the leading eigenvalue $\lambda(q)$ of its integral kernel $K$. In particular, the definition of the scaled cumulant generating function, Eq.~\eqref{eq:scaledcumul}, implies that 
\begin{equation}\label{eq:tilting}
\psi^{(\tau)}(q)=\ln \lambda(q) .
\end{equation}
Combining Eq.~\eqref{eq:generalizedEL} and Eq.~\eqref{eq:tilting} we obtain $2\lambda(-\Lambda)=1$. Calling $\phi(\tau)$ the eigenvector associated with this eigenvalue, we express this result as
\begin{equation}\label{eq:lebow}
\phi(\tau)=2\int d\tau'\, e^{-\Lambda \tau'} p(\tau|\tau') \phi(\tau'). 
\end{equation}
Equation~\eqref{eq:lebow} is a classic result for the Markovian case \cite{powell1956growth,lebowitz1974theory,lin2020single}. The eigenvector $\phi(\tau)$ can be interpreted as the distribution of division times that one would measure over the whole tree, including cells that have not divided yet at time $T$ \cite{levien2020interplay}.

In conclusion, in this paper we derived the generalized Euler-Lotka equation~\eqref{eq:generalizedEL}. This equation describes the growth rate of populations where cell divisions occur in a correlated way. We obtained this result by means of a result in queuing theory \cite{glynn1994large} that was recently applied in stochastic thermodynamics \cite{gingrich2017fundamental} and to study enzymes replicating information \cite{chiuchiu2019error}. We have demonstrated that our result can be easily applied to lineage data. A comparison with the prediction of the traditional Euler-Lotka equation permits to quantitatively assess the impact of correlations on the population growth rate. Due to these properties, we expect the generalized Euler-Lotka equation to become a useful tool to analyze lineages in experimental population dynamics.

\begin{acknowledgments}
I thank Deepak Bhat and Anzhelika Koldaeva for discussions. I am grateful to Arthur Genthon, David Lacoste, Luca Peliti, and two anonymous referees for feedback on the manuscript.
\end{acknowledgments}

\bigskip

\appendix

\section{Estimate of the exponential growth rate $\Lambda$}\label{app:lambda}

We briefly detail the numerical estimation of the exponential growth rate $\Lambda$, as defined by Eq.~\eqref{eq:lambda}, in the adder model by Campos et al. \cite{campos2014constant}. We perform stochastic simulations of the adder model starting with one individual. To ensure that the size and division time of the initial individual are compatible with the steady-state values, we first evolve a single lineage for a time $T_{\mathrm{trans}}=10^5$ and assign to the initial individual the values of $\alpha$, $l$, and reproduction time of the individual representing the lineage at the final time. 

\begin{figure}[htb]
\begin{center}
\includegraphics[width=9cm]{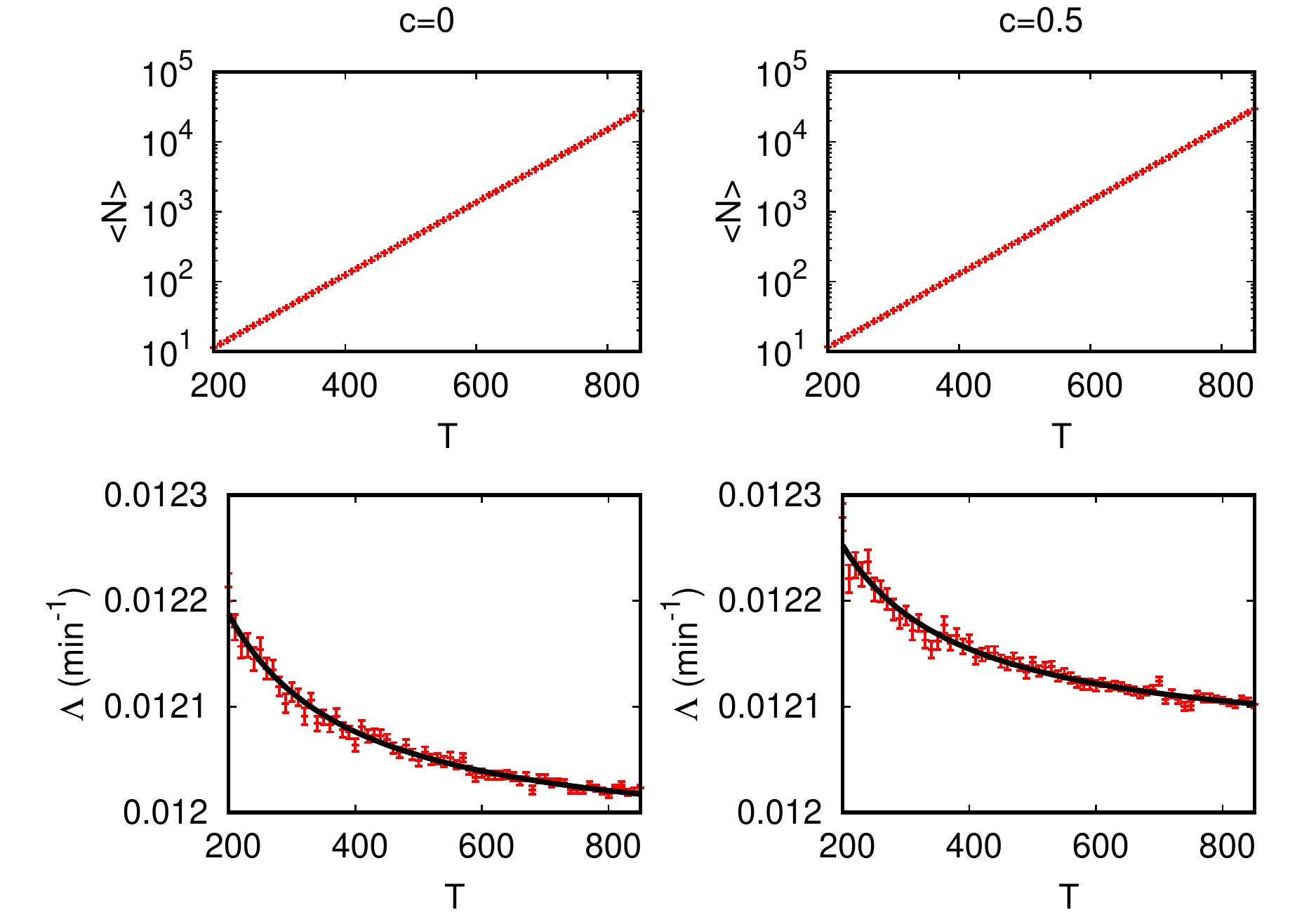}
\caption{Direct estimate of the growth rate $\Lambda$ in the model by Campos et al. \cite{campos2014constant}. (top panels) Average number of individuals $\langle N(T)\rangle$ as a function of time $T$. Parameters are the same as in the Main Text.  (bottom panels) Finite time exponential growth rate $\Lambda(T)$ as a function of $T$, see Eq.~\eqref{eq:lambda_finite}. The black line is a fit according to Eq.~\eqref{eq:lambda_finite2}. Fitted parameters are $\Lambda=0.0119649 ~\mathrm{min}^{-1}$, $C=0.0445346$ for $c=0$, and $\Lambda=0.0120565 ~\mathrm{min}^{-1}$, $C=0.0391681$ for $c=0.5$. In both cases, the relative error on $\Lambda$ from the fit is on the order of $10^{-4}$.
\label{fig:supp}}
\end{center}
\end{figure}

We then reset the time to zero and evolve an entire population tree from this initial individual up to a given time $T$. This simulation is performed using a next reaction scheme. Individuals are stored into a list. Each individual is characterized by its growth rate $\alpha$, its length $l$, and the time $T_{\ind}$ of its next cell division. At each simulation step, we choose the individual with the lowest $T_{\ind}$, advance the time to $T_{\ind}$, and create two new individuals from the chosen one by the division rules detailed in the Main Text. This procedure is repeated until a time $T$ to obtain the number of individual $N(T)$. We average over $10^4$ realizations and compute the mean $\langle N(T)\rangle$ and variance $\sigma^2_{N(T)}$ from these realizations. The behavior of $\langle N(T)\rangle$ as a function of $T$ is shown in Fig.~\ref{fig:supp}, top panels for $c=0$ and $c=0.5$.

We then estimate the finite-time version of the exponential growth rate as 
\begin{equation}\label{eq:lambda_finite}
\Lambda(T)=\frac{1}{T}\ln \langle N(T)\rangle .
\end{equation}
By running simulations for different values of $T$, we find that $\Lambda(T)$ tends to decrease with $T$, and that these finite-time effects are well described by the empirical expression
\begin{equation}\label{eq:lambda_finite2}
\Lambda(T)\approx \Lambda+\frac{C}{T},
\end{equation}
see Fig.~\ref{fig:supp}, bottom panels. We extrapolate to infinite time by fitting Eq.~\eqref{eq:lambda_finite2} to our numerical results. This procedure leads to the estimates of $\Lambda$ shown as black lines in Fig.~2 of the Main Text.

\begin{figure}[htb]
\begin{center}
\includegraphics[width=9cm]{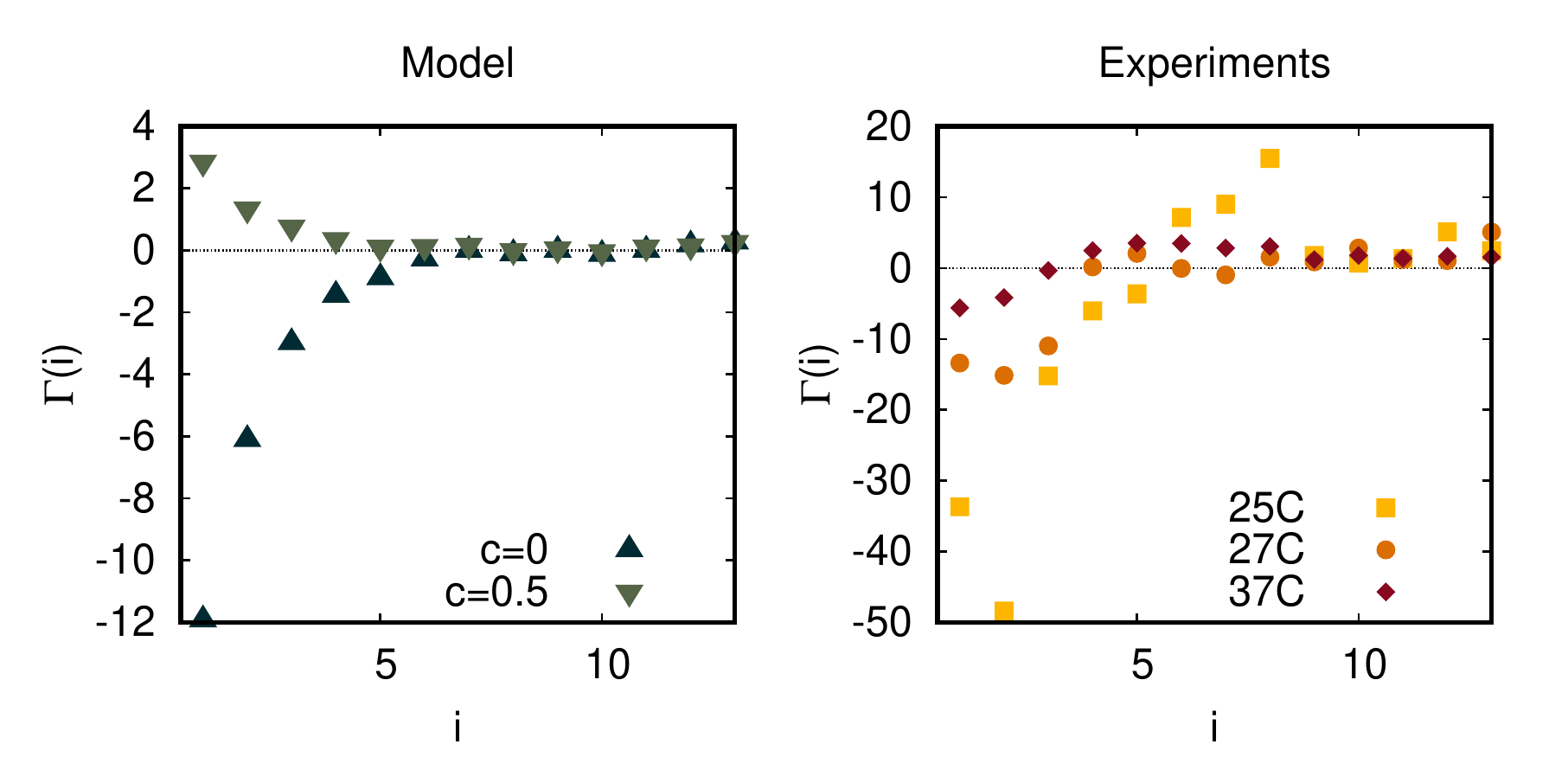}
\caption{Correlation functions of the division times for (left) the model by Campos et al.  \cite{campos2014constant} with same parameters as in the Main text, and (right) experiments by Tanouchi et al. \cite{tanouchi2015noisy,tanouchi2017long} at different temperatures.
\label{fig:corr}}
\end{center}
\end{figure}

\section{Correlation functions}\label{app:corr}

In this Appendix we present the correlation function of cell division times across lineages, defined by

\begin{equation}
\Gamma(i)=\langle \tau_j \tau_{j+i}\rangle -\langle \tau\rangle^2,
\end{equation}
where the average is taken over $j$ and over independent lineages. The behavior of $\Gamma(i)$ for the model by Campos et al. \cite{campos2014constant} and for the experimental data by Tanouchi et al. \cite{tanouchi2015noisy,tanouchi2017long} is shown in Fig. ~\ref{fig:corr}.

\bibliography{eulerlotka}

\begin{thebibliography}{25}%
\makeatletter
\providecommand \@ifxundefined [1]{%
 \@ifx{#1\undefined}
}%
\providecommand \@ifnum [1]{%
 \ifnum #1\expandafter \@firstoftwo
 \else \expandafter \@secondoftwo
 \fi
}%
\providecommand \@ifx [1]{%
 \ifx #1\expandafter \@firstoftwo
 \else \expandafter \@secondoftwo
 \fi
}%
\providecommand \natexlab [1]{#1}%
\providecommand \enquote  [1]{``#1''}%
\providecommand \bibnamefont  [1]{#1}%
\providecommand \bibfnamefont [1]{#1}%
\providecommand \citenamefont [1]{#1}%
\providecommand \href@noop [0]{\@secondoftwo}%
\providecommand \href [0]{\begingroup \@sanitize@url \@href}%
\providecommand \@href[1]{\@@startlink{#1}\@@href}%
\providecommand \@@href[1]{\endgroup#1\@@endlink}%
\providecommand \@sanitize@url [0]{\catcode `\\12\catcode `\$12\catcode
  `\&12\catcode `\#12\catcode `\^12\catcode `\_12\catcode `\%12\relax}%
\providecommand \@@startlink[1]{}%
\providecommand \@@endlink[0]{}%
\providecommand \url  [0]{\begingroup\@sanitize@url \@url }%
\providecommand \@url [1]{\endgroup\@href {#1}{\urlprefix }}%
\providecommand \urlprefix  [0]{URL }%
\providecommand \Eprint [0]{\href }%
\providecommand \doibase [0]{http://dx.doi.org/}%
\providecommand \selectlanguage [0]{\@gobble}%
\providecommand \bibinfo  [0]{\@secondoftwo}%
\providecommand \bibfield  [0]{\@secondoftwo}%
\providecommand \translation [1]{[#1]}%
\providecommand \BibitemOpen [0]{}%
\providecommand \bibitemStop [0]{}%
\providecommand \bibitemNoStop [0]{.\EOS\space}%
\providecommand \EOS [0]{\spacefactor3000\relax}%
\providecommand \BibitemShut  [1]{\csname bibitem#1\endcsname}%
\let\auto@bib@innerbib\@empty
\bibitem [{\citenamefont {Hashimoto}\ \emph {et~al.}(2016)\citenamefont
  {Hashimoto}, \citenamefont {Nozoe}, \citenamefont {Nakaoka}, \citenamefont
  {Okura}, \citenamefont {Akiyoshi}, \citenamefont {Kaneko}, \citenamefont
  {Kussell},\ and\ \citenamefont {Wakamoto}}]{hashimoto2016noise}%
  \BibitemOpen
  \bibfield  {author} {\bibinfo {author} {\bibfnamefont {M.}~\bibnamefont
  {Hashimoto}}, \bibinfo {author} {\bibfnamefont {T.}~\bibnamefont {Nozoe}},
  \bibinfo {author} {\bibfnamefont {H.}~\bibnamefont {Nakaoka}}, \bibinfo
  {author} {\bibfnamefont {R.}~\bibnamefont {Okura}}, \bibinfo {author}
  {\bibfnamefont {S.}~\bibnamefont {Akiyoshi}}, \bibinfo {author}
  {\bibfnamefont {K.}~\bibnamefont {Kaneko}}, \bibinfo {author} {\bibfnamefont
  {E.}~\bibnamefont {Kussell}}, \ and\ \bibinfo {author} {\bibfnamefont
  {Y.}~\bibnamefont {Wakamoto}},\ }\href@noop {} {\bibfield  {journal}
  {\bibinfo  {journal} {Proceedings of the National Academy of Sciences}\
  }\textbf {\bibinfo {volume} {113}},\ \bibinfo {pages} {3251} (\bibinfo {year}
  {2016})}\BibitemShut {NoStop}%
\bibitem [{\citenamefont {Grassly}\ and\ \citenamefont
  {Fraser}(2008)}]{grassly2008mathematical}%
  \BibitemOpen
  \bibfield  {author} {\bibinfo {author} {\bibfnamefont {N.~C.}\ \bibnamefont
  {Grassly}}\ and\ \bibinfo {author} {\bibfnamefont {C.}~\bibnamefont
  {Fraser}},\ }\href@noop {} {\bibfield  {journal} {\bibinfo  {journal} {Nature
  Reviews Microbiology}\ }\textbf {\bibinfo {volume} {6}},\ \bibinfo {pages}
  {477} (\bibinfo {year} {2008})}\BibitemShut {NoStop}%
\bibitem [{\citenamefont {Wang}\ \emph {et~al.}(2010)\citenamefont {Wang},
  \citenamefont {Robert}, \citenamefont {Pelletier}, \citenamefont {Dang},
  \citenamefont {Taddei}, \citenamefont {Wright},\ and\ \citenamefont
  {Jun}}]{wang2010robust}%
  \BibitemOpen
  \bibfield  {author} {\bibinfo {author} {\bibfnamefont {P.}~\bibnamefont
  {Wang}}, \bibinfo {author} {\bibfnamefont {L.}~\bibnamefont {Robert}},
  \bibinfo {author} {\bibfnamefont {J.}~\bibnamefont {Pelletier}}, \bibinfo
  {author} {\bibfnamefont {W.~L.}\ \bibnamefont {Dang}}, \bibinfo {author}
  {\bibfnamefont {F.}~\bibnamefont {Taddei}}, \bibinfo {author} {\bibfnamefont
  {A.}~\bibnamefont {Wright}}, \ and\ \bibinfo {author} {\bibfnamefont
  {S.}~\bibnamefont {Jun}},\ }\href@noop {} {\bibfield  {journal} {\bibinfo
  {journal} {Current biology}\ }\textbf {\bibinfo {volume} {20}},\ \bibinfo
  {pages} {1099} (\bibinfo {year} {2010})}\BibitemShut {NoStop}%
\bibitem [{\citenamefont {Campos}\ \emph {et~al.}(2014)\citenamefont {Campos},
  \citenamefont {Surovtsev}, \citenamefont {Kato}, \citenamefont {Paintdakhi},
  \citenamefont {Beltran}, \citenamefont {Ebmeier},\ and\ \citenamefont
  {Jacobs-Wagner}}]{campos2014constant}%
  \BibitemOpen
  \bibfield  {author} {\bibinfo {author} {\bibfnamefont {M.}~\bibnamefont
  {Campos}}, \bibinfo {author} {\bibfnamefont {I.~V.}\ \bibnamefont
  {Surovtsev}}, \bibinfo {author} {\bibfnamefont {S.}~\bibnamefont {Kato}},
  \bibinfo {author} {\bibfnamefont {A.}~\bibnamefont {Paintdakhi}}, \bibinfo
  {author} {\bibfnamefont {B.}~\bibnamefont {Beltran}}, \bibinfo {author}
  {\bibfnamefont {S.~E.}\ \bibnamefont {Ebmeier}}, \ and\ \bibinfo {author}
  {\bibfnamefont {C.}~\bibnamefont {Jacobs-Wagner}},\ }\href@noop {} {\bibfield
   {journal} {\bibinfo  {journal} {Cell}\ }\textbf {\bibinfo {volume} {159}},\
  \bibinfo {pages} {1433} (\bibinfo {year} {2014})}\BibitemShut {NoStop}%
\bibitem [{\citenamefont {Wallden}\ \emph {et~al.}(2016)\citenamefont
  {Wallden}, \citenamefont {Fange}, \citenamefont {Lundius}, \citenamefont
  {Baltekin},\ and\ \citenamefont {Elf}}]{wallden2016synchronization}%
  \BibitemOpen
  \bibfield  {author} {\bibinfo {author} {\bibfnamefont {M.}~\bibnamefont
  {Wallden}}, \bibinfo {author} {\bibfnamefont {D.}~\bibnamefont {Fange}},
  \bibinfo {author} {\bibfnamefont {E.~G.}\ \bibnamefont {Lundius}}, \bibinfo
  {author} {\bibfnamefont {{\"O}.}~\bibnamefont {Baltekin}}, \ and\ \bibinfo
  {author} {\bibfnamefont {J.}~\bibnamefont {Elf}},\ }\href@noop {} {\bibfield
  {journal} {\bibinfo  {journal} {Cell}\ }\textbf {\bibinfo {volume} {166}},\
  \bibinfo {pages} {729} (\bibinfo {year} {2016})}\BibitemShut {NoStop}%
\bibitem [{\citenamefont {Amir}(2014)}]{amir2014cell}%
  \BibitemOpen
  \bibfield  {author} {\bibinfo {author} {\bibfnamefont {A.}~\bibnamefont
  {Amir}},\ }\href@noop {} {\bibfield  {journal} {\bibinfo  {journal} {Physical
  review letters}\ }\textbf {\bibinfo {volume} {112}},\ \bibinfo {pages}
  {208102} (\bibinfo {year} {2014})}\BibitemShut {NoStop}%
\bibitem [{\citenamefont {Taheri-Araghi}\ \emph {et~al.}(2015)\citenamefont
  {Taheri-Araghi}, \citenamefont {Bradde}, \citenamefont {Sauls}, \citenamefont
  {Hill}, \citenamefont {Levin}, \citenamefont {Paulsson}, \citenamefont
  {Vergassola},\ and\ \citenamefont {Jun}}]{taheri2015cell}%
  \BibitemOpen
  \bibfield  {author} {\bibinfo {author} {\bibfnamefont {S.}~\bibnamefont
  {Taheri-Araghi}}, \bibinfo {author} {\bibfnamefont {S.}~\bibnamefont
  {Bradde}}, \bibinfo {author} {\bibfnamefont {J.~T.}\ \bibnamefont {Sauls}},
  \bibinfo {author} {\bibfnamefont {N.~S.}\ \bibnamefont {Hill}}, \bibinfo
  {author} {\bibfnamefont {P.~A.}\ \bibnamefont {Levin}}, \bibinfo {author}
  {\bibfnamefont {J.}~\bibnamefont {Paulsson}}, \bibinfo {author}
  {\bibfnamefont {M.}~\bibnamefont {Vergassola}}, \ and\ \bibinfo {author}
  {\bibfnamefont {S.}~\bibnamefont {Jun}},\ }\href@noop {} {\bibfield
  {journal} {\bibinfo  {journal} {Current biology}\ }\textbf {\bibinfo {volume}
  {25}},\ \bibinfo {pages} {385} (\bibinfo {year} {2015})}\BibitemShut
  {NoStop}%
\bibitem [{\citenamefont {Powell}(1956)}]{powell1956growth}%
  \BibitemOpen
  \bibfield  {author} {\bibinfo {author} {\bibfnamefont {E.}~\bibnamefont
  {Powell}},\ }\href@noop {} {\bibfield  {journal} {\bibinfo  {journal}
  {Microbiology}\ }\textbf {\bibinfo {volume} {15}},\ \bibinfo {pages} {492}
  (\bibinfo {year} {1956})}\BibitemShut {NoStop}%
\bibitem [{\citenamefont {Lebowitz}\ and\ \citenamefont
  {Rubinow}(1974)}]{lebowitz1974theory}%
  \BibitemOpen
  \bibfield  {author} {\bibinfo {author} {\bibfnamefont {J.~L.}\ \bibnamefont
  {Lebowitz}}\ and\ \bibinfo {author} {\bibfnamefont {S.}~\bibnamefont
  {Rubinow}},\ }\href@noop {} {\bibfield  {journal} {\bibinfo  {journal}
  {Journal of Mathematical Biology}\ }\textbf {\bibinfo {volume} {1}},\
  \bibinfo {pages} {17} (\bibinfo {year} {1974})}\BibitemShut {NoStop}%
\bibitem [{\citenamefont {Lin}\ and\ \citenamefont
  {Amir}(2017)}]{lin2017effects}%
  \BibitemOpen
  \bibfield  {author} {\bibinfo {author} {\bibfnamefont {J.}~\bibnamefont
  {Lin}}\ and\ \bibinfo {author} {\bibfnamefont {A.}~\bibnamefont {Amir}},\
  }\href@noop {} {\bibfield  {journal} {\bibinfo  {journal} {Cell systems}\
  }\textbf {\bibinfo {volume} {5}},\ \bibinfo {pages} {358} (\bibinfo {year}
  {2017})}\BibitemShut {NoStop}%
\bibitem [{\citenamefont {Levien}\ \emph
  {et~al.}(2020{\natexlab{a}})\citenamefont {Levien}, \citenamefont {Kondev},\
  and\ \citenamefont {Amir}}]{levien2020interplay}%
  \BibitemOpen
  \bibfield  {author} {\bibinfo {author} {\bibfnamefont {E.}~\bibnamefont
  {Levien}}, \bibinfo {author} {\bibfnamefont {J.}~\bibnamefont {Kondev}}, \
  and\ \bibinfo {author} {\bibfnamefont {A.}~\bibnamefont {Amir}},\ }\href@noop
  {} {\bibfield  {journal} {\bibinfo  {journal} {Journal of the Royal Society
  Interface}\ }\textbf {\bibinfo {volume} {17}},\ \bibinfo {pages} {20190827}
  (\bibinfo {year} {2020}{\natexlab{a}})}\BibitemShut {NoStop}%
\bibitem [{\citenamefont {Kobayashi}\ and\ \citenamefont
  {Sughiyama}(2015)}]{kobayashi2015fluctuation}%
  \BibitemOpen
  \bibfield  {author} {\bibinfo {author} {\bibfnamefont {T.~J.}\ \bibnamefont
  {Kobayashi}}\ and\ \bibinfo {author} {\bibfnamefont {Y.}~\bibnamefont
  {Sughiyama}},\ }\href@noop {} {\bibfield  {journal} {\bibinfo  {journal}
  {Physical review letters}\ }\textbf {\bibinfo {volume} {115}},\ \bibinfo
  {pages} {238102} (\bibinfo {year} {2015})}\BibitemShut {NoStop}%
\bibitem [{\citenamefont {Garc{\'\i}a-Garc{\'\i}a}\ \emph
  {et~al.}(2019)\citenamefont {Garc{\'\i}a-Garc{\'\i}a}, \citenamefont
  {Genthon},\ and\ \citenamefont {Lacoste}}]{garcia2019linking}%
  \BibitemOpen
  \bibfield  {author} {\bibinfo {author} {\bibfnamefont {R.}~\bibnamefont
  {Garc{\'\i}a-Garc{\'\i}a}}, \bibinfo {author} {\bibfnamefont
  {A.}~\bibnamefont {Genthon}}, \ and\ \bibinfo {author} {\bibfnamefont
  {D.}~\bibnamefont {Lacoste}},\ }\href@noop {} {\bibfield  {journal} {\bibinfo
   {journal} {Physical Review E}\ }\textbf {\bibinfo {volume} {99}},\ \bibinfo
  {pages} {042413} (\bibinfo {year} {2019})}\BibitemShut {NoStop}%
\bibitem [{\citenamefont {Genthon}\ and\ \citenamefont
  {Lacoste}(2020{\natexlab{a}})}]{genthon2020fluctuation}%
  \BibitemOpen
  \bibfield  {author} {\bibinfo {author} {\bibfnamefont {A.}~\bibnamefont
  {Genthon}}\ and\ \bibinfo {author} {\bibfnamefont {D.}~\bibnamefont
  {Lacoste}},\ }\href@noop {} {\bibfield  {journal} {\bibinfo  {journal}
  {Scientific Reports}\ }\textbf {\bibinfo {volume} {10}},\ \bibinfo {pages}
  {1} (\bibinfo {year} {2020}{\natexlab{a}})}\BibitemShut {NoStop}%
\bibitem [{\citenamefont {Genthon}\ and\ \citenamefont
  {Lacoste}(2020{\natexlab{b}})}]{genthon2020universal}%
  \BibitemOpen
  \bibfield  {author} {\bibinfo {author} {\bibfnamefont {A.}~\bibnamefont
  {Genthon}}\ and\ \bibinfo {author} {\bibfnamefont {D.}~\bibnamefont
  {Lacoste}},\ }\href@noop {} {\bibfield  {journal} {\bibinfo  {journal} {arXiv
  preprint arXiv:2012.02734}\ } (\bibinfo {year}
  {2020}{\natexlab{b}})}\BibitemShut {NoStop}%
\bibitem [{\citenamefont {Levien}\ \emph
  {et~al.}(2020{\natexlab{b}})\citenamefont {Levien}, \citenamefont
  {GrandPre},\ and\ \citenamefont {Amir}}]{levien2020large}%
  \BibitemOpen
  \bibfield  {author} {\bibinfo {author} {\bibfnamefont {E.}~\bibnamefont
  {Levien}}, \bibinfo {author} {\bibfnamefont {T.}~\bibnamefont {GrandPre}}, \
  and\ \bibinfo {author} {\bibfnamefont {A.}~\bibnamefont {Amir}},\ }\href@noop
  {} {\bibfield  {journal} {\bibinfo  {journal} {Physical Review Letters}\ ,\
  \bibinfo {pages} {048102}} (\bibinfo {year}
  {2020}{\natexlab{b}})}\BibitemShut {NoStop}%
\bibitem [{\citenamefont {Touchette}(2009)}]{touchette2009large}%
  \BibitemOpen
  \bibfield  {author} {\bibinfo {author} {\bibfnamefont {H.}~\bibnamefont
  {Touchette}},\ }\href@noop {} {\bibfield  {journal} {\bibinfo  {journal}
  {Physics Reports}\ }\textbf {\bibinfo {volume} {478}},\ \bibinfo {pages} {1}
  (\bibinfo {year} {2009})}\BibitemShut {NoStop}%
\bibitem [{\citenamefont {Nozoe}\ \emph {et~al.}(2017)\citenamefont {Nozoe},
  \citenamefont {Kussell},\ and\ \citenamefont
  {Wakamoto}}]{nozoe2017inferring}%
  \BibitemOpen
  \bibfield  {author} {\bibinfo {author} {\bibfnamefont {T.}~\bibnamefont
  {Nozoe}}, \bibinfo {author} {\bibfnamefont {E.}~\bibnamefont {Kussell}}, \
  and\ \bibinfo {author} {\bibfnamefont {Y.}~\bibnamefont {Wakamoto}},\
  }\href@noop {} {\bibfield  {journal} {\bibinfo  {journal} {PLoS genetics}\
  }\textbf {\bibinfo {volume} {13}},\ \bibinfo {pages} {e1006653} (\bibinfo
  {year} {2017})}\BibitemShut {NoStop}%
\bibitem [{\citenamefont {Jafarpour}\ \emph {et~al.}(2018)\citenamefont
  {Jafarpour}, \citenamefont {Wright}, \citenamefont {Gudjonson}, \citenamefont
  {Riebling}, \citenamefont {Dawson}, \citenamefont {Lo}, \citenamefont
  {Fiebig}, \citenamefont {Crosson}, \citenamefont {Dinner},\ and\
  \citenamefont {Iyer-Biswas}}]{jafarpour2018bridging}%
  \BibitemOpen
  \bibfield  {author} {\bibinfo {author} {\bibfnamefont {F.}~\bibnamefont
  {Jafarpour}}, \bibinfo {author} {\bibfnamefont {C.~S.}\ \bibnamefont
  {Wright}}, \bibinfo {author} {\bibfnamefont {H.}~\bibnamefont {Gudjonson}},
  \bibinfo {author} {\bibfnamefont {J.}~\bibnamefont {Riebling}}, \bibinfo
  {author} {\bibfnamefont {E.}~\bibnamefont {Dawson}}, \bibinfo {author}
  {\bibfnamefont {K.}~\bibnamefont {Lo}}, \bibinfo {author} {\bibfnamefont
  {A.}~\bibnamefont {Fiebig}}, \bibinfo {author} {\bibfnamefont
  {S.}~\bibnamefont {Crosson}}, \bibinfo {author} {\bibfnamefont {A.~R.}\
  \bibnamefont {Dinner}}, \ and\ \bibinfo {author} {\bibfnamefont
  {S.}~\bibnamefont {Iyer-Biswas}},\ }\href@noop {} {\bibfield  {journal}
  {\bibinfo  {journal} {Physical Review X}\ }\textbf {\bibinfo {volume} {8}},\
  \bibinfo {pages} {021007} (\bibinfo {year} {2018})}\BibitemShut {NoStop}%
\bibitem [{\citenamefont {Glynn}\ and\ \citenamefont
  {Whitt}(1994)}]{glynn1994large}%
  \BibitemOpen
  \bibfield  {author} {\bibinfo {author} {\bibfnamefont {P.~W.}\ \bibnamefont
  {Glynn}}\ and\ \bibinfo {author} {\bibfnamefont {W.}~\bibnamefont {Whitt}},\
  }\href@noop {} {\bibfield  {journal} {\bibinfo  {journal} {Queueing Systems}\
  }\textbf {\bibinfo {volume} {17}},\ \bibinfo {pages} {107} (\bibinfo {year}
  {1994})}\BibitemShut {NoStop}%
\bibitem [{\citenamefont {Tanouchi}\ \emph {et~al.}(2015)\citenamefont
  {Tanouchi}, \citenamefont {Pai}, \citenamefont {Park}, \citenamefont {Huang},
  \citenamefont {Stamatov}, \citenamefont {Buchler},\ and\ \citenamefont
  {You}}]{tanouchi2015noisy}%
  \BibitemOpen
  \bibfield  {author} {\bibinfo {author} {\bibfnamefont {Y.}~\bibnamefont
  {Tanouchi}}, \bibinfo {author} {\bibfnamefont {A.}~\bibnamefont {Pai}},
  \bibinfo {author} {\bibfnamefont {H.}~\bibnamefont {Park}}, \bibinfo {author}
  {\bibfnamefont {S.}~\bibnamefont {Huang}}, \bibinfo {author} {\bibfnamefont
  {R.}~\bibnamefont {Stamatov}}, \bibinfo {author} {\bibfnamefont {N.~E.}\
  \bibnamefont {Buchler}}, \ and\ \bibinfo {author} {\bibfnamefont
  {L.}~\bibnamefont {You}},\ }\href@noop {} {\bibfield  {journal} {\bibinfo
  {journal} {Nature}\ }\textbf {\bibinfo {volume} {523}},\ \bibinfo {pages}
  {357} (\bibinfo {year} {2015})}\BibitemShut {NoStop}%
\bibitem [{\citenamefont {Tanouchi}\ \emph {et~al.}(2017)\citenamefont
  {Tanouchi}, \citenamefont {Pai}, \citenamefont {Park}, \citenamefont {Huang},
  \citenamefont {Buchler},\ and\ \citenamefont {You}}]{tanouchi2017long}%
  \BibitemOpen
  \bibfield  {author} {\bibinfo {author} {\bibfnamefont {Y.}~\bibnamefont
  {Tanouchi}}, \bibinfo {author} {\bibfnamefont {A.}~\bibnamefont {Pai}},
  \bibinfo {author} {\bibfnamefont {H.}~\bibnamefont {Park}}, \bibinfo {author}
  {\bibfnamefont {S.}~\bibnamefont {Huang}}, \bibinfo {author} {\bibfnamefont
  {N.~E.}\ \bibnamefont {Buchler}}, \ and\ \bibinfo {author} {\bibfnamefont
  {L.}~\bibnamefont {You}},\ }\href@noop {} {\bibfield  {journal} {\bibinfo
  {journal} {Scientific data}\ }\textbf {\bibinfo {volume} {4}},\ \bibinfo
  {pages} {1} (\bibinfo {year} {2017})}\BibitemShut {NoStop}%
\bibitem [{\citenamefont {Lin}\ and\ \citenamefont
  {Amir}(2020)}]{lin2020single}%
  \BibitemOpen
  \bibfield  {author} {\bibinfo {author} {\bibfnamefont {J.}~\bibnamefont
  {Lin}}\ and\ \bibinfo {author} {\bibfnamefont {A.}~\bibnamefont {Amir}},\
  }\href@noop {} {\bibfield  {journal} {\bibinfo  {journal} {Physical Review
  E}\ }\textbf {\bibinfo {volume} {101}},\ \bibinfo {pages} {012401} (\bibinfo
  {year} {2020})}\BibitemShut {NoStop}%
\bibitem [{\citenamefont {Gingrich}\ and\ \citenamefont
  {Horowitz}(2017)}]{gingrich2017fundamental}%
  \BibitemOpen
  \bibfield  {author} {\bibinfo {author} {\bibfnamefont {T.~R.}\ \bibnamefont
  {Gingrich}}\ and\ \bibinfo {author} {\bibfnamefont {J.~M.}\ \bibnamefont
  {Horowitz}},\ }\href@noop {} {\bibfield  {journal} {\bibinfo  {journal}
  {Physical review letters}\ }\textbf {\bibinfo {volume} {119}},\ \bibinfo
  {pages} {170601} (\bibinfo {year} {2017})}\BibitemShut {NoStop}%
\bibitem [{\citenamefont {Chiuchi{\'u}}\ \emph {et~al.}(2019)\citenamefont
  {Chiuchi{\'u}}, \citenamefont {Tu},\ and\ \citenamefont
  {Pigolotti}}]{chiuchiu2019error}%
  \BibitemOpen
  \bibfield  {author} {\bibinfo {author} {\bibfnamefont {D.}~\bibnamefont
  {Chiuchi{\'u}}}, \bibinfo {author} {\bibfnamefont {Y.}~\bibnamefont {Tu}}, \
  and\ \bibinfo {author} {\bibfnamefont {S.}~\bibnamefont {Pigolotti}},\
  }\href@noop {} {\bibfield  {journal} {\bibinfo  {journal} {Physical review
  letters}\ }\textbf {\bibinfo {volume} {123}},\ \bibinfo {pages} {038101}
  (\bibinfo {year} {2019})}\BibitemShut {NoStop}%
\end{thebibliography}%

\end{document}